# Nitrogen investigation by SIMS in two wide band-gap semiconductors: Diamond and Silicon Carbide

Marie-Amandine PINAULT-THAURY* and François JOMARD

Université Paris-Saclay, UVSQ, CNRS, GEMaC, 78000 Versailles, FRANCE

*corresponding author:  marie-amandine.pinault-thaury@uvsq.fr

**Abstract**

Diamond and Silicon Carbide (SiC) are promising wide band-gap semiconductors for power electronics, SiC being more mature especially in term of large wafer size (that has reached 200 mm quite recently). Nitrogen impurities are often used in both materials for different purpose: increase the diamond growth rate or induce n-type conductivity in SiC. The determination of the nitrogen content by secondary ion mass spectrometry (SIMS) is a difficult task mainly because nitrogen is an atmospheric element for which direct monitoring of $N^{\pm}$ ions give no or a weak signal. With our standard diamond SIMS conditions, we investigate 12C14N- secondary ions under cesium primary ions by applying high mass resolution settings. Nitrogen depth-profiling of diamond and SiC (multi-) layers is then possible over several micrometer thick over reasonable time analysis duration. In a simple way and without notably modifying our usual analysis process, we found a nitrogen detection limit of $2 \times 10^{17}$ at/cm$^3$ in diamond and $5 \times 10^{15}$ at/cm$^3$ in SiC.

**Introduction**

As other wide bandgap semiconductors, diamond and silicon carbide (SiC) exhibits remarkable physical properties that make them suitable for power components [1, 2]. Compared to silicon, electronic components based on diamond or SiC allow a considerable improvement of the device properties. For SiC, nitrogen doping increases the amount of current flowing through a transistor. Although this doping technology appears mature, the understanding of some synthesis steps remains unsolved. A better knowledge of the structural properties of nitrogen-doped SiC is therefore essential [3-5]. For diamond, the use of nitrogen during its synthesis allows to increase the growth rate [6-8]. However, to avoid deleterious effects on diamond properties, the nitrogen content has to be as low as possible. Indeed, the determination of low doping levels of nitrogen is then essential for both materials.

Light elements, hydrogen, carbon, nitrogen and oxygen, are known as making up the atmosphere. They are present in a lot of materials and in all vacuum chambers. At low content (<1 at%), their presence in materials is difficult to detect. Among the microanalysis techniques, only Secondary Ion Mass Spectrometry (SIMS) can detect them. For SIMS analysis the material is placed under ultra-high vacuum (up to a few $10^{-10}$ mbar) and sputtered with a beam of primary ions. The secondary ions resulting from the sputtering are collected and sorted by a mass spectrometer.

In the semiconductor field, SIMS analysis is usually performed to determine the depth distribution and concentration of dopants but also to study the incorporation of elements resulting from the synthesis process. In order to achieve optimum sensitivity, elements with a low potential for first ionization are detected as positive secondary ions whereas elements with an electro-negative feature are detected as negative secondary ions.

For 20 years, we perform intensive research on diamond in terms of homoepitaxial growth and chemical analysis by SIMS. In this work, we propose to apply and adjust our method of analyzing nitrogen in diamond to silicon carbide. We will emphasis on the nitrogen detection limit, that is to say the smallest measurable nitrogen concent, in these two materials.

**Experimental details**

**Dynamic SIMS technique**

The SIMS technique consists in sputtering a solid sample, placed under vacuum, by a focused primary ion beam which is rastered over a square area. This destructive analytical technique can be applied to any type of solid material that can be kept under vacuum. The sputtering by primary ions of a few keV energy induces the ejection of particles from the sample. Part of the ejected particles are ionized (positively or negatively). The sputtered secondary ions are analyzed with a mass spectrometer. The SIMS technique provides a unique combination of extremely high sensitivity for all chemical elements of the periodic table with a very low background. This allows high dynamic range (up to more than 5 decades) and low detection limits (down to ppb level, depending on the SIMS settings, the analyzed element and the studied material).

With dynamic SIMS, the bulk composition and the in-depth distribution of trace elements are investigated with a good depth resolution (typically lower than 10 nm). This technique gives information about the elemental and isotopic composition through the material thickness (up to a few tens of μm). The secondary ion yields will vary greatly according to the chemical environment and the sputtering conditions (ion, energy, angle). Even if this adds complexity to the quantitative aspect of the technique, SIMS is recognized as the most sensitive elemental and isotopic analysis technique.

**Standard SIMS diamond conditions**

The SIMS analysis were performed with dynamic magnetic sector SIMS (IMS7f-CAMECA). For diamond homoepitaxy, SIMS is usually performed in order to measure the depth distribution of dopants (boron or phosphorus) and contaminants (as hydrogen) in the films. In routine analysis, the raster size is 150x150 μm². The analyzed zone is restricted to a diameter of 33 μm to limit the crater edge effects. The depth of the resulting SIMS crater is measured with a Dektak8 step-meter

The analysis are usually performed at low mass resolution, M/ΔM ~ 400, to have a maximum sensitivity. As the SIMS technique is mostly used in diamond to detect impurities, measurements are performed with parameters allowing high sensitivities. Namely the $Cs^+/M^-$ configuration is classically employed: positive primary ions with a $Cs^+$ source and detection of negative secondary ions of mass M. The energy of the $Cs^+$ primary beam is set to 10 keV. Secondary ions are detected in the negative mode by biasing the sample to −5000 V, leading to an interaction energy of the primary ions of 15 keV and an incidence angle of 23° with respect to the normal of the sample. By using a primary intensity in the range of 40 to 60 nA, leading to a sputtering rate of around 0.25 to 0.4 nm/s in diamond matrix, such diamond SIMS conditions do not require time-consuming adjustment.

The detected masses are usually the one coming from the matrix (as $^{12}C$) and the one coming from the homoepilayer (as the dopant impurities $^{11}B$ or $^{31}P$ [9, 10]).

**Detection of nitrogen (low mass resolution)**

Diamond for electronics requires plasma assisted chemical vapor deposition synthesis that allows low impurity contamination. Moreover, diamond homoepilayers are often grown on diamond substrates that contain a high nitrogen level (in the range of $10^{19}$ at/cm$^3$). If thick layers (>20 μm) can be estimated by weight or comparator techniques, the thickness of thin diamond homoepilayers is only accessible by SIMS thanks to the difference in nitrogen content between the layer and the substrat. Unfortunately, with zero electron affinity, nitrogen give no negative ion $^{14}N^-$. The most efficient way to detect nitrogen is then to collect a molecular ion consisting of a nitrogen atom and a constituent atom of the matrix (ex: detection of $^{14}N^{28}Si^-$ ions in silicon [11]). Knowing that carbon is the only matrix element in diamond, the $^{12}C^{14}N^-$ molecular ion is use to detect nitrogen with a good sensitivity (see Fig.1). As a result, we use this molecular ion to analyze nitrogen in diamond and SiC. The corresponding detected mass, M, is close to 26 u.m.a..

**Determination of the nitrogen content: high mass resolution for $^{12}C^{14}N^-$ fine tuning**

With low mass resolution SIMS analysis, the 26 u.m.a signal contains $^{12}C^{14}N^-$ (26.003074 u.m.a.) with all mass interferences as $^{13}C_2^-$ (26. 00671 u.m.a.). The $^{12}C^{14}N^-$ mass is then not clearly distinguisible from neighboring masses (see Fig.2(a)). Fortunatly, our apparatus allows to work at high resolution in mass, that is to say to separate $^{12}C^{14}N^-$ ions from its neighboring ions. The adjustement consists of slightly closing the entrance and exit slits of the spectrometer, just enough to get M/ΔM close to 7 500 (see Fig.2(b)). It is worth to note that the use of HMR settings reduces the signal intensities of the SIMS analysis by few decades.

In this work, we have applied our standard SIMS diamond conditions together with HMR settings on two different samples: a low nitrogen doped diamond homoepilayer ($6 \times 10^{17}$ at/cm$^3$) and a lightly doped SiC layer ($2 \times 10^{16}$ at/cm$^3$). We have detected the $^{13}C_2^-$ ion as matrix element that has an in-depth constant level, and $^{12}C^{14}N^-$, the element of interest.

**Results and discussion**

**Light elements: a specific case**

The analysis of the so-called atmospheric elements presents an additional difficulty. These elements constitute the residual vacuum of the analysis chamber. Under these conditions, the measured signal can then come from the material but also from background. To limit background contribution, a specific care is taken to reach a vacuum limit of $\sim 5 \times 10^{-10}$ mbar in the analysis chamber. The following paragraph describes the protocol used to determine the limit of detection in both materials.

**Raster size method**

We have focused our attention on the determination of the detection limit, DL, reached in such SIMS conditions. For that purpose, we have chosen the raster size method (another method consists in sputtering with high rates to reduce the background contribution). It consists in a step by step reduction of the raster size. Each time the raster size is decreased, there is a simultaneous increase of the primary ion density, Jp (ratio between the primary intensity, and the raster size), and the intensity of the detected secondary ions. Fig. 3(a) shows the results obtained on the SiC sample by varying the raster size from 250x250 µm² down to 100x100 µm². The same procedure has been performed on the diamond sample from 150x150 µm² down to 75x75 µm² (not shown here).

In Fig.3(b), the mean intensity of the $^{12}C^{14}N^-$ signal over each raster size, I($^{12}C^{14}N^-$), is reported for each primary density, Jp, of each raster size for diamond and SiC samples. In such a plot, the slope of the line corresponds to the fraction of signal actually coming from the material and which is proportional to Jp. The y-intercept corresponds to the constant signal fraction coming from the background. It is this latter value which sets the detection limit under given analysis conditions. Therfore, DL is given here by the y-intercept of the linear fit of the data I($^{12}C^{14}N^-$) = f(Jp). In our case, the intercepts are ~267 cps for diamond and ~8 cps for SiC.

**Quantification of DL**

Quantitative SIMS analysis requires standard materials to determine a relative sensitivity factor (RSF) values. In the case of diamond, we use an implanted standard with a known dose of nitrogen while for SiC, we use a bulk SiC with a known nitrogen content. The formula allowing the quantification of DL is given by Eq.1:

$$[DL] = DL*RSF/I_{matrix} \qquad (1)$$

where $I_{matrix}$ is the mean intensity of the matrix element detected during the analysis (expressed in cps), DL (in cps) is extracted from the raster size method and RSF (at/cm$^3$) is given by the SIMS analysis of the standard materials. The uncertainty of the DL quantification is evaluated to 20%. Table 1 resumes the values obtained from our SIMS analysis by using the $^{13}C_2^-$ ions as matrix element.

The [DL] value in diamond (2x10$^{17}$ at/cm$^3$) is quite high compare to SiC (5x10$^{15}$ at/cm$^3$). As the best DL in SIMS should correspond to the detection of 0 or 1 cps, it is necessary to use a sample presenting low enough content of the element of interest but still high enough to be detectable. That is why the raster size method is usually performed on lightly doped samples (namely in the range of 10$^{16}$ at/cm$^3$). With higher content, the determination of the DL is no longer precise due to high secondary intensities. Therefore, the extracted DL value may suffer of a high uncertainty.

In our case, the diamond sample used contains 30 time more nitrogen than the SiC sample. As a result, the mean intensities of $^{12}C^{14}N^-$ measured with the raster size method are much more important in the diamond sample (> 500 cps) than in the SiC sample (< 50 cps). Then the error induced in the DL determination is more important for the diamond sample. This could partially explain the factor 20 between the [DL] values of diamond and SiC. The use of a diamond sample with lightly nitrogen doped level (~1x10$^{16}$ at/cm$^3$) should lead to a more accurate value of DL. However, such samples are hard to obtain because the precise control of nitrogen incorporation during diamond growth is a challenge at such low doping level.

In the literature, there are just a few reported studies on nitrogen DL in SiC [12-17] by detecting different nitrogen-matrix ions (as SiN [13]). Most of this studies are performed with higher primary ion intensities (from 80 nA up to 400 nA) than in our work. The common point is that the lower DL achievable are time consuming and cannot be guaranteed at all the times. Wang et al. [14, 15] use SIMS settings that are limited to bulk and thick layers of SiC (> 2 μm) with constant nitrogen content. Indeed, small raster size (50x50 μm² and lower) often leads to un-flat bottom crater that is unwanted for analyzing samples with impurity concentration variation [16]. Moreover, lower raster size induces faster sputter rate and less data points, unsuitable for (ultra-) thin multilayer SiC samples [13]. For Ya Der et al. [17], low DL requires great hardware overheads and retains the transmission of the
analytical channel. As a result, it is necessary to evaluate the relevance of our SIMS conditions on multi-layer SiC sample.

For this reason, we have applied our standard SIMS diamond conditions with HMR settings to another diamond sample and the previous SiC sample through several micro-meters. By adjusting the raster size and the primary ion intensity to have reasonable sputtering rates, we obtain the depth profiles of nitrogen presented in Fig.4. The profiles of 3 μm in diamond and 8 μm in SiC can be done in only ~2h and ~3h, respectively, keeping a low DL, especially in SiC. We observe that the diamond homoepilayer presents a nitrogen signal in the range of the diamond [DL] value. For SiC sample, the buried layer underneath the thick lightly doped layer is also at the SiC [DL] value. Such depth-profiling validates our SIMS settings for the analysis of SiC multi-layers with low nitrogen DL.

**Summary**

In this work, we have investigated nitrogen detection by SIMS. We have determined the detection limit, DL, of nitrogen in diamond and SiC. We used our standard diamond SIMS conditions with high mass resolution (HMR) settings that require few adjustments, are reliable and not time-consuming. We found a nitrogen DL value of ~$5 \times 10^{15}$ at/cm$^3$ for SiC, nearly 20 time lower than in diamond. Such value allows the determination of nitrogen in lightly nitrogen doped SiC layers (> $5 \times 10^{15}$ at/cm$^3$). In the future, we will continue to investigate nitrogen DL in both wide-band gap materials by adjusting HMR SIMS settings to try to reach lower values.

**Acknowledgements**

The authors thank the "region Ile de France", the National Center for Scientific Research (CNRS), the university of Versailles (UVSQ), the Paris-Saclay center of the French Alternative Energies and Atomic Energy Commission (CEA), the EDF company and the École Polytechnique (PolytechX) for having financing the IMS7f-CAMECA apparatus that is an open facility located in the GEMaC laboratory.

**Table caption**

Table 1: Values of RSF, $I(^{13}C_2^-)$, DL and [DL] for both wide band-gap semiconductors obtained using diamond SIMS analysis with HMR.

**Figure caption**

Figure 1: $^{12}$C, $^{11}$B and $^{12}$C$^{14}$N profiles in a diamond delta-doped structure grown on a diamond substrate. The intensity of $^{12}$C (matrix element) is intense and constant because their is no change in material between the structure and the substrate (all in diamond). The intensity of $^{11}$B (only present in the delta layers) is not changing at the interface because there is no boron in the substrate and up to the first delta layer. The interface between the structure (that does not contain nitrogen) and the substrate (that contains nitrogen) is only visualised thanks to the step marked by the $^{12}$C$^{14}$N signal.

Figure 2: Mass spectra around 26 u.m.a. at (a) low mass resolution where $^{12}$C$^{14}$N$^-$ ions are not separated from neighboring ions and (b) high mass resolution where $^{12}$C$^{14}$N$^-$ ions are clearly separate from neighboring ions (as $^{13}$C$_2^-$).

Figure 3: (a) Raster size method applied to a lightly nitrogen doped SiC layer (the primary intensity is fixed at 40nA). The step by step decrease of the raster size from 250x250 µm² down to 100x100 µm² induces an increase of the $^{12}$C$^{14}$N$^-$ intensity. (b) Mean intensity of the $^{12}$C$^{14}$N$^-$ signal, I($^{12}$C$^{14}$N$^-$), in function of the primary ion density, Jp, extracted from the raster size method (upper graph for diamond and lower graph for SiC). The dashed lines are the linear fit of the experimental data. The intercept of the linear fit gives the detection limit, DL.

Figure 4: HMR nitrogen depth profiles of the diamond and SiC samples. The raster size and the primary ion intensity were adjusted to reach a sputtering rate around 0.55 ± 0.15 nm/s (150x150 µm² and 60 nA for diamond; 100x100 µm² and 40 nA for SiC). The dashed areas correspond to the nitrogen DL with an accuracy of 20%.

| Material | Diamond | SiC |
|---|---|---|
| RSF in at/cm$^3$ | 3.30x10$^{18}$ | 1.34x10$^{18}$ |
| I($^{13}$C$^-$) in cps | ~4400 | ~2100 |
| DL in cps | ~267 | ~8 |
| [DL] in at/cm$^3$ | ~2x10$^{17}$ | ~5x10$^{15}$ |

Table 1

Pinault-Thaury et al.

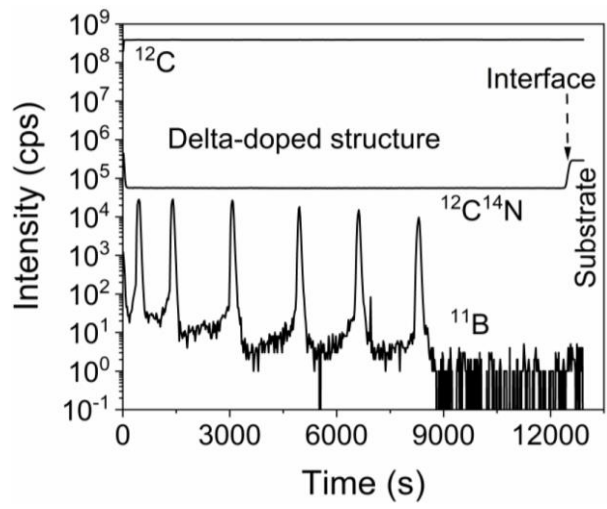

Figure 1

Pinault-Thaury et al.

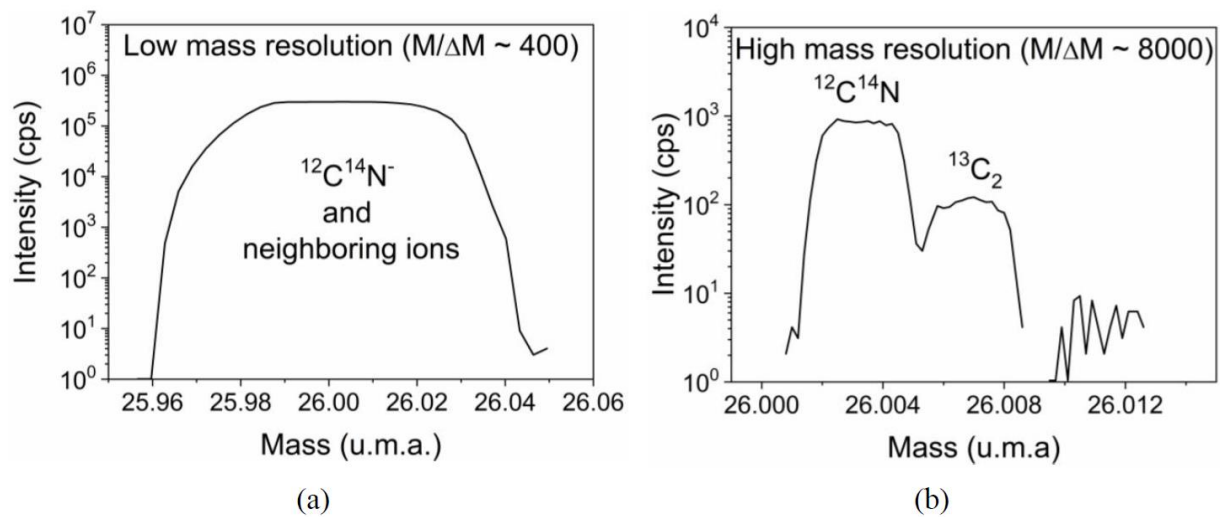

Figure 2

Pinault-Thaury et al.

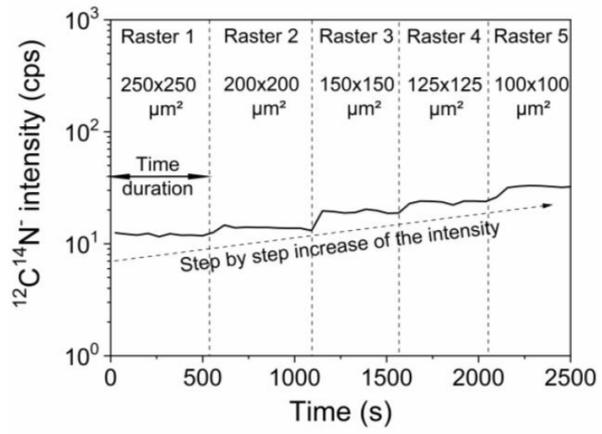   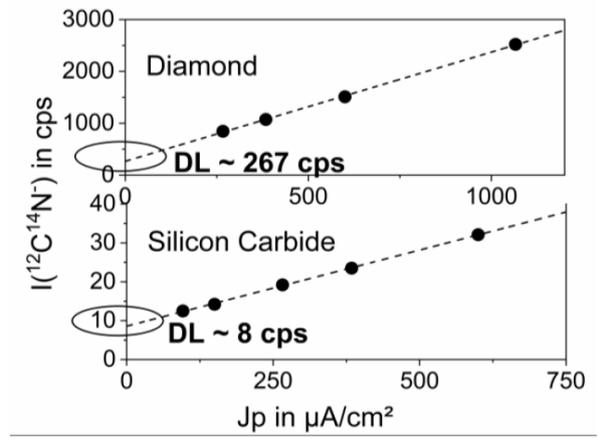

(a)                              (b)

Figure 3

Pinault-Thaury et al.

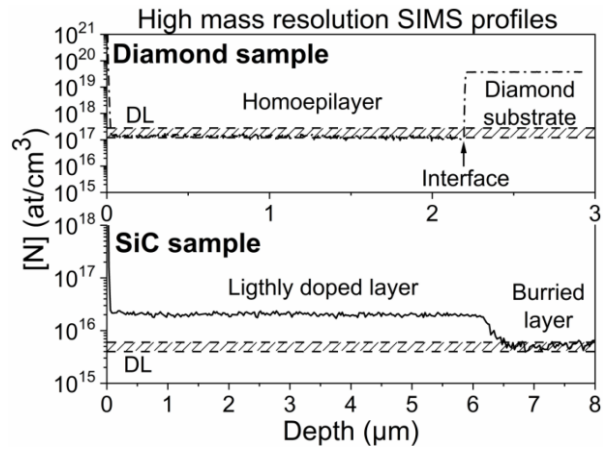

Figure 4

Pinault-Thaury et al.